\documentclass[prb,twocolumn,superscriptaddress,showpacs]{revtex4-1}
\usepackage{graphicx}
\usepackage{amsmath}

\begin{document}

\title{Parameterizing the surface free energy and excess adsorption of a hard-sphere fluid at a planar hard wall}
\author{Ruslan L. Davidchack}
\affiliation{Department of Mathematics, University of Leicester, Leicester, LE1 7RH, United Kingdom}
\author{Brian B. Laird}
\affiliation{Department of Chemistry, University of Kansas, Lawrence, KS 66045, United States}
\author{Roland Roth}
\affiliation{Institut f\"{u}r Theoretische Physik, Universit\"{a}t T\"{u}bingen, D-72074 T\"{u}bingen, Germany}
\date{\today}
\begin{abstract}
The inhomogeneous structure of a fluid at a wall can be characterized
in several ways. Within a thermodynamic description the surface free
energy $\gamma$ and the excess adsorption $\Gamma$ are of central
importance. For theoretical studies closed expression of $\gamma$ and
$\Gamma$ can be very valuable; however, even for a well-studied model
system such as a hard-sphere fluid at a planar hard wall, the accuracy
of  existing expressions for $\gamma$ and $\Gamma$, compared to precise
computer simulation data, can still be improved. Here, we compare
several known expressions for $\gamma$ and $\Gamma$ to the most precise computer simulation data. While good agreement is generally found at low to intermediate
fluid densities, the existing parameterizations show significant deviation at high density. In this work, we propose 
new parameterizations for $\gamma$ and $\Gamma$ that agree with the simulation data within statistical error over the entire fluid
density range.
\end{abstract}
\keywords{surface thermodynamics, molecular-dynamics simulation, adsorption}
\maketitle

\section{Introduction}

The hard-sphere fluid at a hard wall is a useful reference model for the solid-liquid interface between chemically dissimilar materials. Its simple, but non-trivial, nature has made it a standard reference model to test theories of inhomogeneous fluids, such as integral equation theories and classical density-functional theories. As such, there have been a large number of simulation efforts to study the detailed thermodynamics and structure of this system, both to provide data for the testing of theoretical methods and to provide insight into the generic phenomenology of solid-liquid interfaces. Of particular interest in such studies is the surface free energy, $\gamma$, which measures the work required to create a unit area of interface, and the excess adsorption, $\Gamma$, which measures the number of particles in the interfacial region relative to that in a region of equal volume in the bulk. In terms of the single-particle density profile, $\rho(z)$, the excess adsorption is given by
\begin{equation}
\Gamma = \int_0^\infty [\rho(z) - \rho]\; dz
\end{equation}
where $\rho$ is the bulk fluid density and $z$ is the Cartesian coordinate normal to the surface.

The excess adsorption and the surface free energy are related through the Gibbs adsorption equation
\begin{equation} 
\Gamma = -\left( \frac{\partial \gamma}{\partial \mu} \right )_{T}
\label{eq:gibbs}
\end{equation}
where $\mu$ is the chemical potential. A more convenient relationship for use in molecular simulation studies is one derived from the Gibbs-Cahn procedure\cite{Laird10} for the excess volume, $v$:
\begin{equation}
v = \left (\frac{\partial \gamma}{\partial P} \right )_{T}
\label{eq:gibbs-cahn_v}
\end{equation}
The excess adsorption is directly related to the excess volume by the relation\cite{Laird10}
\begin{equation}
 \Gamma = -\rho\,v
\label{eq:excess_v}
\end{equation}
For the hard-sphere/hard-wall system, we can express both $\gamma$ and $\Gamma$ in dimensionless form:  $\gamma^* = \beta \gamma \sigma^2$ and $\Gamma^* = \Gamma \sigma^2$, where $\beta = 1/kT$ and $\sigma$ is the hard-sphere diameter. In what follows, we will drop the $^*$ and assume that all quantities are in dimensionless form. 

For the hard-sphere fluid/hard-wall system, the values of $\gamma$ and $\Gamma$ are dependent upon the choice of the reference point for measuring distance between the wall and the fluid spheres.  In this work, we will adopt the "edge-centered convention", where the coordinate of the center of a fluid sphere in contact with the wall is $z = \sigma/2$.  In contract, a number of other studies - especially many of the early works - use the "sphere-centered" convention with $z = 0$ begin the coordinate of the center of such a sphere.  These two conventions give different values of the system volume and other characteristics, but the relationship between them is easy to establish. If we denote the interfacial free energy for a system using the edge-centered and sphere-centered conventions as $\gamma$ and $\bar{\gamma}$, respectively, then we have
\begin{equation}
\gamma = \bar{\gamma} + \frac{P}{2}
\label{eq:gammabar}
\end{equation}
where $P$ is the bulk fluid pressure. 
Correspondingly, if we denote the edge-centered and sphere-centered values of the excess adsorption as $\Gamma$ and $\bar{\Gamma}$, respectively, we have
\begin{equation}
\Gamma = \bar{\Gamma} - \frac{\rho}{2}
\end{equation}

Over the past four decades, there have been a number of simulation studies focused on the calculation of $\gamma$ and $\Gamma$ for the hard-sphere/hard-wall system. One of the earliest is that of Henderson and van Swol\cite{Henderson84}, who calculate $\gamma$ using a mechanical definition of the surface tension via the Kirkwood-Buff equation\cite{Kirkwood49}
\begin{equation}
\gamma = \int_{-\infty}^{\infty} [ p_n (z) - p_t(z) ]\; dz
\label{eq:Kirkwood-Buff}
\end{equation}
where $p_n$ and $p_t$ are the normal and transverse components of the pressure tensor and $z$ is the direction normal to the wall. In that early work, $\gamma$ (and also $\Gamma$) are calculated at relatively few values of the packing fraction $\eta = \pi \rho \sigma^3/6$. The statistical uncertainties in these calculation are also quite high, due both to convergence issues inherent in Eq.~\ref{eq:Kirkwood-Buff} and computational power at the time. More recently, de Miguel and Jackson using an improved Kirkwood-Buff algorithm calculated $\gamma$ with higher statistical precision than the results of Ref.~\onlinecite{Henderson84}, but the values at the highest packing fractions are in disagreement with most other later studies. Heni and L\"owen\cite{Heni99} used a more accurate thermodynamic integration technique with square-barrier and triangular cleaving potentials to determine $\gamma$. While considerably improved over the Henderson and van Swol results, the relative statistical error in $\gamma$ at the highest packing fraction studied (near fluid-solid coexistence) was still high (nearly 10\%). This thermodynamic integration method was improved upon by Fortini and Dijkstra\cite{Fortini06} with significant improvement in the overall statistical error. The most precise  published measurements to date are those of Laird and Davidchack\cite{Laird10} and recent calculations by Yang, et al.\cite{JHYang13}. The former were calculated using the Gibbs-Duhem integration technique\cite{Frolov09a,Laird09,Laird10} from data for the excess volume (from which $\Gamma$ can be easily calculated), whereas the latter utilizes a grand-canonical transition matrix Monte Carlo simulation method. Recently, we have refined the calculations in Ref.~\onlinecite{Laird10} to be more precise. These new results are given in the Supplemental Information.\cite{supp} These simulation data are identical to those in Ref.~\onlinecite{Laird10} within the original statistical estimates, but are considerably more precise. 

For many theoretical and practical applications using the hard-sphere/hard-wall as a reference model, it is useful to have an accurate parametrized form of the wall surface
tension $\gamma(\eta)$ and of excess adsorption $\Gamma(\eta)$ that
has a simple form and accurately accounts for the simulation
data. In this work, we critically review previous theoretical and empirical expressions for $\gamma$ and $\Gamma$ and propose our own highly accurate parameterization that fits the most accurate simulation data within the statistical error. 

\section{Theoretical and Empirical Parameterizations for $\gamma$ and $\Gamma$} 
\begin{figure}
\includegraphics[width=0.7\textwidth]{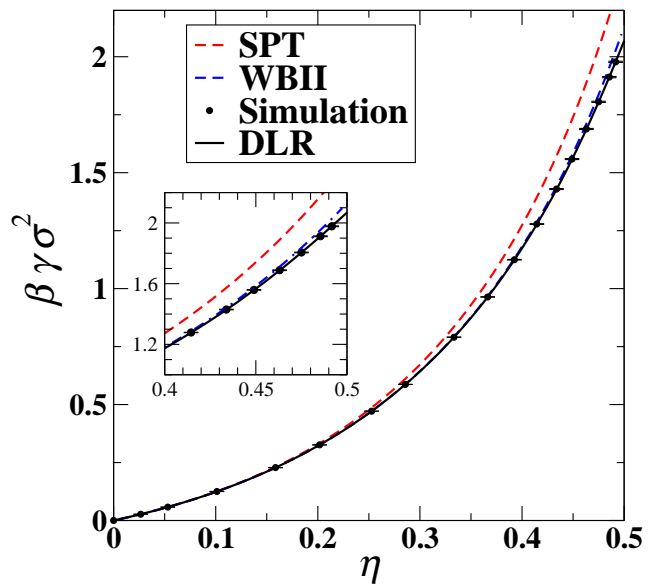}
\caption{\label{fig:comparison} Comparison of the SPT and WBII expressions for the hard-sphere/hard-wall $\gamma$ with the simulation results of Laird and Davidchack.\cite{Laird10,supp} The error bars on the simulation data in this an all subsequent figures represent  95\% confidence estimates.} 
\end{figure}

The earliest functional form for $\gamma$ is that derived from Scaled Particle Theory (SPT):\cite{Reiss60,Barker76}
\begin{equation}
\gamma_\mathrm{SPT} = \frac{3\eta(2+\eta)}{2\pi(1-\eta)^2}
\label{eq:SPT}
\end{equation} 
where $\eta = \pi \rho \sigma^3/6$ is the packing fraction - the fraction of the total volume occupied by the spheres. 
Eq.~\ref{eq:SPT} can be obtained from the expression for $\bar{\gamma}$ given in Refs.~\onlinecite{Reiss60,Barker76} using the SPT expression for the pressure and Eq.~\ref{eq:gammabar}
\begin{equation}
P_\mathrm{SPT} = \frac{6\eta(1+\eta+\eta^2)}{\pi(1-\eta)^3}
\label{eq:P-SPT}
\end{equation} 
Given an expression for $\gamma(\eta)$, the corresponding expression for the excess adsorption can be
obtained using a modified version of Eq.~\ref{eq:excess_v}
\begin{equation}
\Gamma = -\rho \left (\frac{\partial \gamma}{\partial \eta} \right )_T/\left (\frac{\partial P}{\partial \eta} \right )_T
\label{eq:Gamma_partial}
\end{equation}
For the SPT, the expression for $\Gamma$ is quite simple
\begin{equation}
\Gamma_\mathrm{SPT} = -\frac{3\eta (1-\eta)}{\pi(1+2\eta)}
\end{equation}
These SPT expressions work well for very low packing fractions, but exhibit significant deviation from the simulation values at intermediate and high packing fractions - see Fig.~\ref{fig:comparison}.

A more-accurate theoretical expression for $\gamma$ can be derived from density-functional theory (DFT). It has been shown\cite{Bryk03,Koenig04} that an approximation for the surface tension of a hard-sphere fluid at a planar hard wall can be
obtained within a class of DFTs known as Fundamental Measure Theories (FMT).\cite{Rosenfeld89, Roth10} In these theories, the surface free energy can be found from the excess free energy density of a binary bulk
(homogeneous) mixture. For the White Bear Mark II (WBII) version of FMT
one obtains
\begin{equation} \label{eq:WBII}
\gamma_\mathrm{WBII} =
\frac{\eta(2+ 3\eta-2\eta ^2)}{\pi(1-\eta)^2}-
\frac{\ln{(1-\eta )}}{\pi}
\end{equation}
The WBII performs significantly better than the SPT expression and shows significant deviations only at the highest packing fractions - see Fig~\ref{fig:comparison}. This WBII expression will be the starting point of our later parameterization.

The excess adsorption within WBII can be obtained from Eq.~\ref{eq:Gamma_partial} using the Carnahan-Starling equation of state for hard spheres
\begin{equation} 
\beta P_\mathrm{CS} = \frac{6\eta(1 + \eta + \eta^2 - \eta^3)}{\pi(1-\eta)^3}
\label{eq:CS}
\end{equation}
Using $P_\mathrm{CS}$ and Eq.~\ref{eq:WBII} in Eq.~\ref{eq:Gamma_partial} gives
\begin{equation}
\Gamma_\mathrm{WBII} = -\frac{\eta(1-\eta)(3 + 6\eta - 5\eta^2 + 2\eta^3)}{\pi(1+4\eta+4\eta^2-4\eta^3+\eta^4)}
\label{eq:Gamma_WBII}
\end{equation}

The SPT and WBII theoretical expressions for $\gamma$ and $\Gamma$ make convenient starting points for designing empirical expressions to represent the simulation data.  Using the SPT form as a reference, Henderson and Plischke\cite{Henderson85} proposed the following empirical functional form for $\bar{\gamma}$ to fit the molecular-dynamics simulation results of Ref.~\onlinecite{Henderson84}
\begin{equation}
\bar{\gamma}_\mathrm{HP} =  -\frac{9\eta^2(1 + 44\eta/35  -2\eta^2/5)}{2\pi (1-\eta)^3}
\end{equation}
Using Eq.~\ref{eq:gammabar} together with the Carnahan-Starling
equation of state (Eq.~\ref{eq:CS}) gives
\begin{equation}
\gamma_\mathrm{HP} =  \frac{3\eta(2-\eta - 62\eta^2/35-4\eta^3/5)}{2\pi(1-\eta)^3}
\end{equation}
with the corresponding equation for $\Gamma$:
\begin{equation}
\Gamma_\mathrm{HP} = -\frac{3\eta(1+\eta-221\eta^2/70-8\eta^3/5+2\eta^4/5)}{\pi(1+4\eta+4\eta^2-4\eta^3+\eta^4)}
\end{equation}
More recently, Urrutia\cite{Urrutia14}, also starting with the SPT expression, but using the more precise simulation results,\cite{Laird10} suggested a more accurate expression
\begin{equation}
\bar{\gamma}_\mathrm{U} = -\frac{9\eta^2(1+44\eta/35+\eta^2/38 - 3\eta^3 + 3\eta^4 )}{2\pi(1-\eta)^3}
\end{equation}
The corresponding expression for $\gamma$ is
\begin{equation}
\gamma_\mathrm{U} = \frac{3\eta(2-\eta-62\eta^2/35 -  79\eta^3/38+ 9\eta^4  -9\eta^5)}{2\pi(1-\eta)^3}
\end{equation}
This expression has the property that the first three virial coefficients for $\gamma(\eta)$ are exact, as discussed in the next section. Urrutia does not provide a corresponding expression for $\Gamma$, but one can easily derive one using Eq.~\ref{eq:excess_v} and the Carnahan-Starling equation of state
{\tiny
\begin{equation}
\Gamma_\mathrm{U} = -\frac{3\eta(1+\eta-221\eta^2/70 -  79\eta^3/19 + 1789\eta^4/76 -36\eta^5 + 27\eta^6/2)}{\pi(1+4\eta+4\eta^2-4\eta^3+\eta^4)}
\end{equation}
}

\section{Virial Expansions}

Another popular parameterization for any thermodynamic quantity is the so-called virial series, where the quantity of interest is expanded in a Taylor series with respect to the density, pressure or packing fraction. With respect to the packing fraction, the virial expansion for $\gamma$ can be written as
\begin{equation}
\gamma = \sum_{n=1}^{\infty} a_n \eta^n
\label{eq:virial}
\end{equation}
The virial expansion coefficients, $\bar{a}_n$, for $\bar{\gamma}$ can be calculated from those for $\gamma$ using the usual virial expansion coefficients, $B_n$, for the pressure (usually given as an expansion in $\rho$):
\begin{equation}
\frac{\beta P}{\rho} = 1 + \sum_{n=2}^\infty B_n \rho^{n-1}
\end{equation}
The coefficients, $B_n$, are known analytically up to $n = 4$ and have been calculated numerically up to $n = 12$.\cite{Clisby06,Wheatley13} Using Eq.~\ref{eq:gammabar} gives
\begin{equation}
\bar{a}_n = a_n + \frac{B_n}{2}\left (\frac{6}{\pi}\right )^n
\end{equation}
Like the virial coefficients for the pressure,\cite{Hansen06} the virial coefficients $a_n$ can be written as a sum of cluster integrals\cite{Bellemans62} and the first three coefficients of Eq.~\ref{eq:virial} are known analytically.\cite{Stecki78} 
Recently, Yang, et al.\cite{Yang14} used Monte Carlo sampling techniques to evaluate the cluster integrals for $n = 3$ to 7, obtaining approximate estimates for the exact cluster expansion expression for $a_n$. In this work, the virial coefficients, $\bar{a}_n$ for $\bar{\gamma}$ were determined. The virial equation for $\bar{\gamma}$ and the corresponding one for $\Gamma$, truncated at $n=7$, were shown to give very good agreement to their simulation results at low to intermediate packing fractions (up to about $\eta$ = 0.4).  

Each of the theoretical and empirical parameterizations in the previous section can be expanded in a virial series. Table I shows the virial expansion coefficients for $\gamma$ for $n = 1$ to 6. ($n = 7$ is not shown because of the large statistical error in the Monte Carlo estimates of the exact coefficient.) All parameterizations get the exact first two virial coefficients ($3/\pi$ and $15/2\pi$) correctly, but of the parameterizations presented so far, only that of Henderson and Plischke and that of Urrutia also get the correct third virial coefficient ($759/70\pi$).

\begin{table*}[t]
\begin{tabular}{|c||c|c|c|c|c|c|}
\hline
$n$&Exact & SPT &HP& WBII & Urrutia & DLR\\
\hline
1 & $\frac{3}{\pi}$& $\frac{3}{\pi}$&$\frac{3}{\pi}$ &$\frac{3}{\pi}$ &$\frac{3}{\pi}$ &$\frac{3}{\pi}$\\
2 & $\frac{15}{2\pi}$ &$\frac{15}{2\pi}$  & $\frac{15}{2\pi}$ & $\frac{15}{2\pi}$&$\frac{15}{2\pi}$  &$\frac{15}{2\pi}$  \\
3 & $\frac{759}{70\pi}$  = 3.4513.. &$\frac{12}{\pi}$ = 3.8197.. & $\frac{759}{70\pi}$  = 3.4513..& $\frac{31}{3\pi}$  = 3.2892..& $\frac{759}{70 \pi}$ = 3.4513.. & $\frac{158}{15\pi}$ = 3.3528.. \\
4 &3.82(4)  & $\frac{33}{2\pi} = 5.252..$ &$\frac{477}{35\pi}$ = 4.3381..&$\frac{53}{4\pi}$ = 4.2176..&$\frac{26361}{2660 \pi}$ = 3.1545..&$\frac{158}{15\pi}$ = 3.3529..\\
5 & 4.54(8) &$\frac{21}{\pi}$ = 6.6845..&$\frac{111}{7\pi}$ =5.0475 &$\frac{81}{5\pi}$ = 5.1566..& $\frac{48417}{2660\pi}$ = 5.7938..&$\frac{257}{20\pi}$ = 4.0903..\\
6 & 5.5(4) &$\frac{51}{2\pi}$ = 8.1169.. &$\frac{1227}{70\pi}$=5.5795..&$\frac{115}{6\pi}$ = 6.1009...&$\frac{2955}{133\pi}$ = 7.0722.&$\frac{76}{5\pi}$ = 4.8383..\\
\hline
\end{tabular}
\caption{Virial expansion coefficients for the theoretical and empirical parameterizations for $\gamma$ discussed here. The first three "exact" coefficients are known analytically, whereas those for $n =$ 4, 5 and 6 were obtained through Monte Carlo calculation of the cluster integrals involved.\cite{Yang14}}
\end{table*}

\section{A new parameterization}

The Henderson-Plischke and Urrutia parameterizations use as their starting point the SPT expression for $\Gamma$. In this work, we begin with the more accurate WBII expression to build an empirical parameterization. In the WBII, the first two virial coefficients of $\gamma$ agree with the exact values. To preserve this, one can write
\begin{equation}
\gamma_\mathrm{fitting} = \frac{\eta(2 + 3\eta+ a \eta^2 + b\eta^3)}{\pi(1-\eta)^2} - \frac{\ln{(1-\eta)}}{\pi}\;,
\label{eq:fitting}
\end{equation}
where $a$ and $b$ are fitting parameters. For $a = -313/210$, the fitting function also gives the exact third virial coefficient. For this value of $a$, a weighted least-squares fit to the simulation data gives $b = -1.62313$. While this form is a good fit at low to moderate $\eta$, there are still deviations beyond the statistical simulation uncertainties at the very highest packing fractions. 

\begin{figure}
\includegraphics[width=0.45\textwidth]{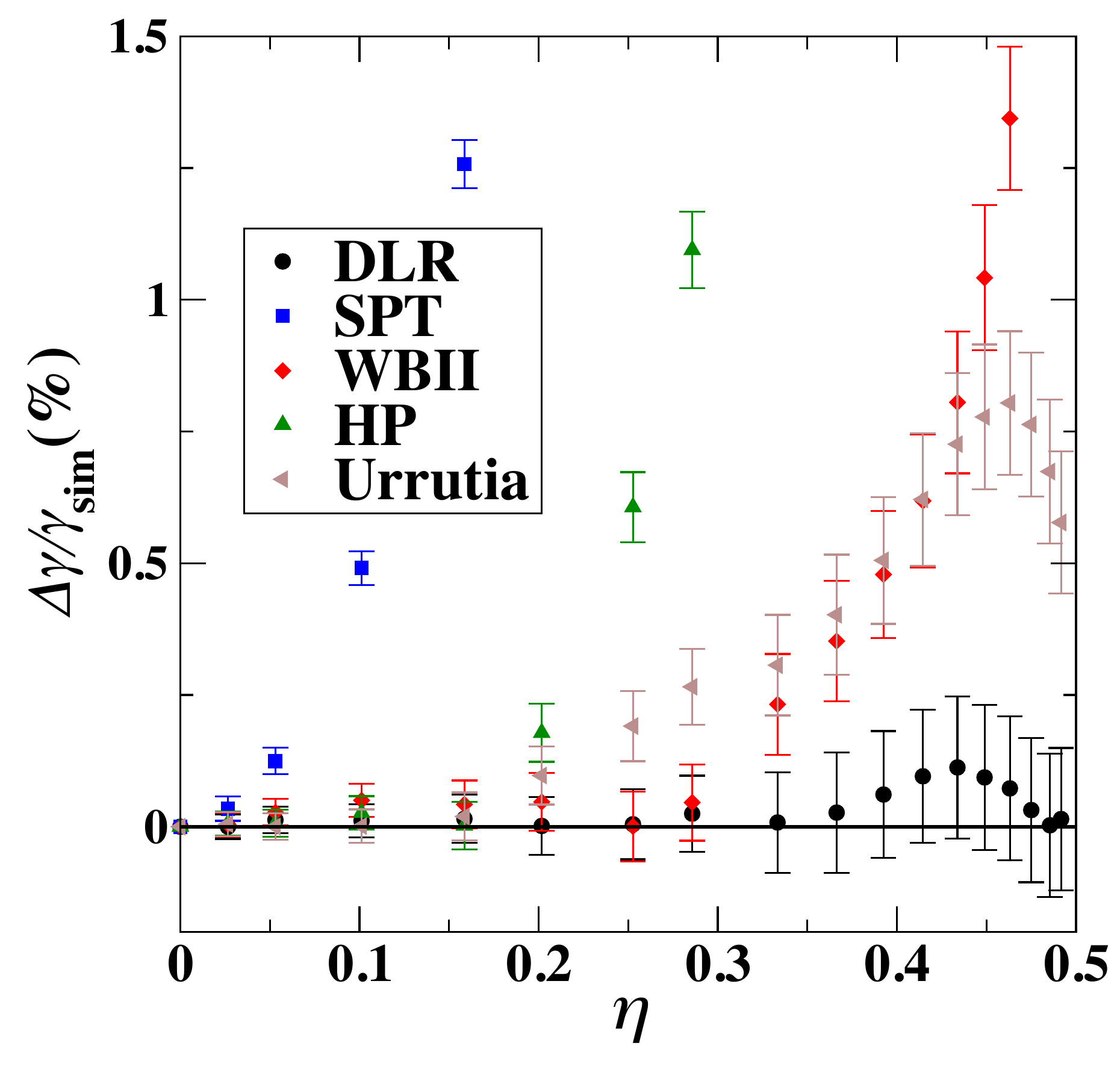}
\caption{\label{fig:gamma_diff} Percentage deviation of the parameterizations for $\gamma$ considered here from the simulation results.\cite{supp}}
\end{figure}

The high packing fraction data can be well fit by introducing a high power term in the numerator of the first term on the right hand side of Eq.~\ref{eq:fitting} - a similar approach was used by  Kolafa, Lab\'{i}k and Malijevsk\'{y} in their development of a highly accurate equation of state for hard spheres.\cite{Kolafa04} The actual high-power exponent ($n$)  used is not too important as long as the term does not significantly affect the value of $\gamma$ at low to intermediate packing fractions - here we use $n = 20$:
\begin{equation}
\gamma_\mathrm{DLR} = \frac{\eta(2 + 3\eta+ a \eta^2 + b\eta^3 + c\eta^{20})}{\pi(1-\eta)^2} - \frac{\ln{(1-\eta)}}{\pi}
\label{eq:fitting2}
\end{equation}
To develop our final parameterization, we note that fixing $a$ to give
the correct third virial coefficient does not lead to an expression that fits the data at intermediate $\eta$ within the 
error bars. Relaxing this condition gives us our final parameterization and the main result of this work. 
Fig.~\ref{fig:gamma_diff} shows the percentage relative deviation of Eq.~\ref{eq:dlr} from the simulation values, along with the deviations for the other parameterizations considered here. Only the new parameterization fits the data within the statistical error over the full range of $\eta$:
{\small
\begin{equation}
\gamma_\mathrm{DLR} = \frac{\eta(2 + 3\eta- 9\eta^2/5 - 4\eta^3/5 - 5\times 10^4 \eta^{20})}{\pi(1-\eta)^2} - \frac{\ln{(1-\eta)}}{\pi} 
\label{eq:dlr}
\end{equation}
}
\begin{figure}
\includegraphics[width=0.45\textwidth]{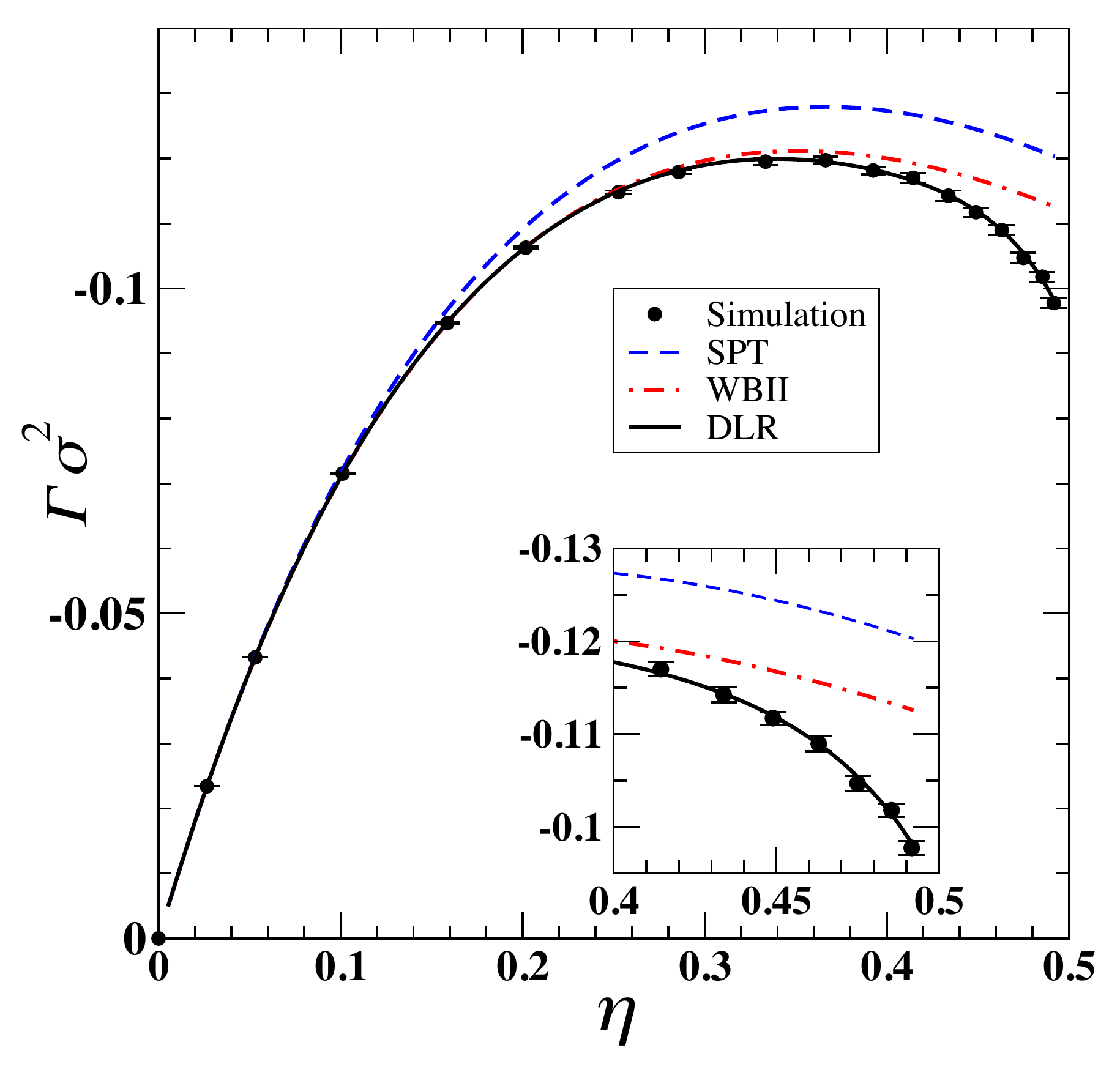}
\caption{\label{fig:adsorption_comparison} Comparison of the SPT, WBII  and DLR parameterizations for the hard-sphere/hard-wall excess adsorption, ($\Gamma$) with the simulation results.\cite{supp} The inset shows the high packing fraction data at higher resolution. }
\end{figure}
Using the Carnahan-Starling equation of state and Eq.~\ref{eq:excess_v}, the corresponding DLR expression for $\Gamma$ is 
{\tiny
\begin{equation}
\Gamma_\mathrm{DLR'} =-\frac{\eta(1-\eta)(15 + 30 \eta -22 \eta^2 - 7\eta^3 + 8\eta^4- 5.75\times10^6\eta^{20} + 4.75\times10^6 \eta^{21})}{5(1 + 4 \eta + 4 \eta^2 - 4 \eta^3 + \eta^4)}
\end{equation}
}
However, this expression does not quite fit the data for $\Gamma$ within the error bars - presumably due to slight inaccuracies in the Carnahan-Starling equation of state; however, slight modification of the high-order terms gives an expression that is accurate over the whole range of the simulation data within statistical error. 
{\tiny
\begin{equation}
\Gamma_\mathrm{DLR} =-\frac{\eta(1-\eta)(15 + 30 \eta -22 \eta^2 - 7\eta^3 + 8\eta^4 - 5.5\times 10^6\eta^{20} + 4.2\times 10^6\eta^{21})}{5(1 + 4 \eta + 4 \eta^2 - 4 \eta^3 + \eta^4)}
\label{eq:DLR_Gamma}
\end{equation}
}
Fig.~\ref{fig:adsorption_comparison} shows the excess adsorption as a function of $\eta$ for the DLR, SPT and WBII expressions (Eq.~\ref{eq:DLR_Gamma}) together with the simulation results\cite{supp}. 

\begin{figure}[h]
\includegraphics[width=0.45\textwidth]{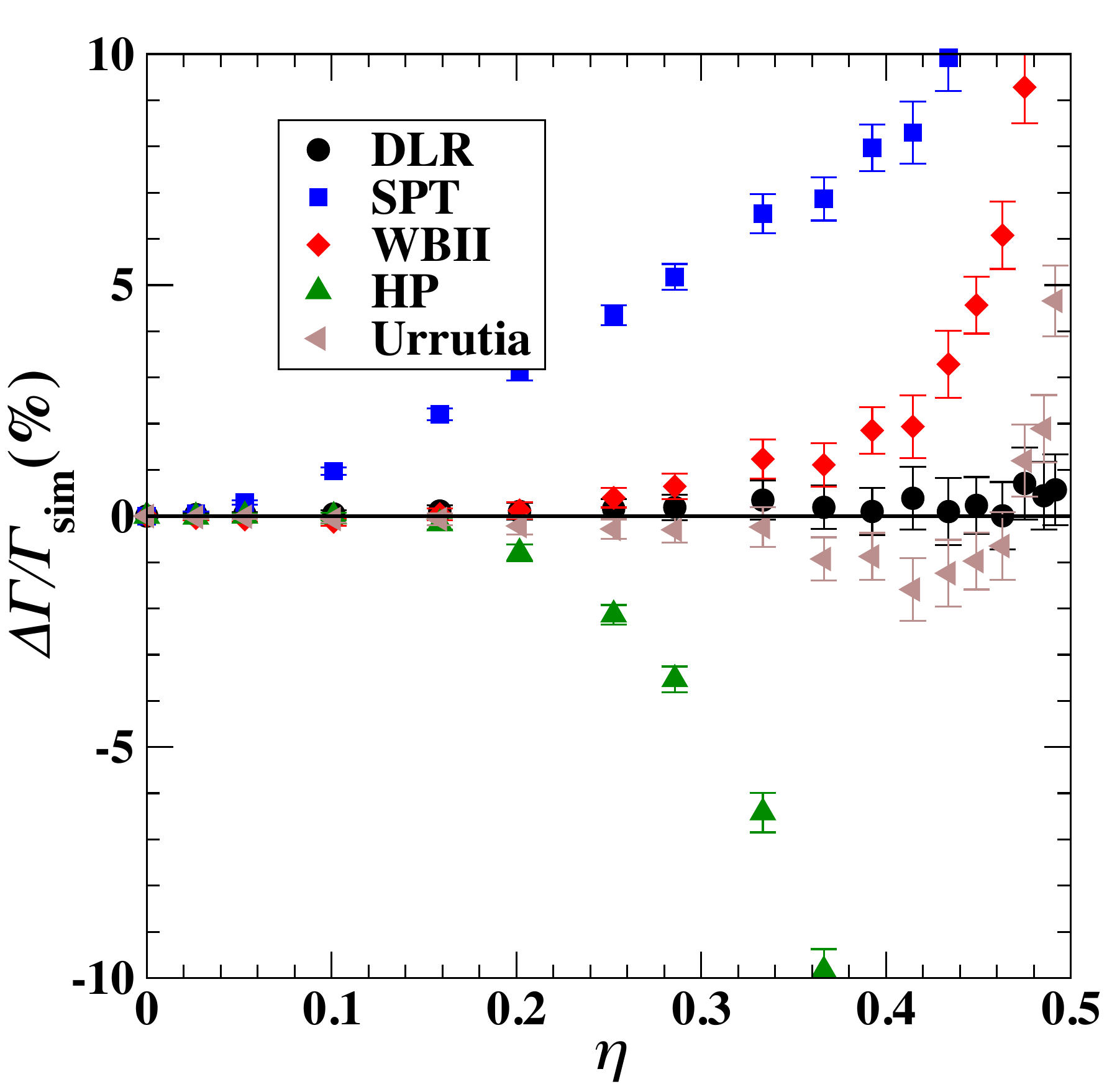}
\caption{\label{fig:adsorption_diff} Percentage deviation from the simulation data\cite{supp} for each of the parameterizations of the adsorption ($\Gamma$) presented here.}
\end{figure}

To compare the adsorption for all of the methods presented here, we plot in Fig.~\ref{fig:adsorption_diff} the percentage deviation from the simulation values as functions of $\eta$ of all of the adsorption expressions presented here. Note that the DLR expression is the only one to capture the simulation data within the statistical error over the entire $\eta$ range studied.

 \section{Summary}

The thermodynamics of an inhomogeneous fluid at a wall can be characterized
by the surface free energy $\gamma$ and the excess adsorption
$\Gamma$, two thermodynamic quantities that are related by
Eq.~\ref{eq:gibbs}, the Gibbs adsorption theorem. For a hard-sphere
fluid at a planar hard wall both $\gamma$ and $\Gamma$ have been
measured very precisely in recent computer simulations. These simulation
data can be employed as benchmark data for theories, such as Scaled Particle Theory (SPT), or
expressions derived from the White Bear Mark II (WBII) density functional theory in certain
limits, or from empirical parameterizations. Each of the expressions for $\gamma$ and $\Gamma$ that
we have tested here perform well at low to intermediate packing
fractions ($\eta$) of the hard-sphere fluid. However, at higher values of $\eta$ close to
freezing there are, however, significant deviations visible in all
existing expressions -- see Figs.~\ref{fig:comparison}--\ref{fig:adsorption_diff}. These
deviations are most prominent for the excess adsorption $\Gamma$.

Because explicit expressions for $\gamma$ and $\Gamma$ for the
hard-sphere hard-wall model system can be useful in theoretical studies
we suggest a new empirical parameterization for these quantities. To
this end we start with the surface free energy $\gamma$.

It is interesting to note that the known virial coefficients for
$\gamma$ are of limited use only in constructing an accurate expression. We employ the functional form
of the surface free energy from WBII with some added fitting
parameters. The key observation, however, is that an additional high power (in
$\eta$) term is required to reproduce the correct behavior of $\gamma$
at high fluid densities. Our parameterization for $\gamma$ is given in
Eq.~\ref{eq:dlr}. 

We obtain the parameterization for the excess adsorption in two
steps. First we employ a modified Gibbs adsorption theorem,
Eq.~\ref{eq:gibbs-cahn_v}, and the Carnahan-Starling equation of state to
find the functional form of $\Gamma$ from that of $\gamma$. Because the
Carnahan-Starling equation of state shows small deviation from
simulation results at very high $\eta$, we can improve the agreement of our
parameterization with the simulation data by slightly adjusting the
higher-order terms, which leads to our final parameterization for $\Gamma$ (Eq.~\ref{eq:DLR_Gamma}).

Both new parameterizations for $\gamma$ and $\Gamma$ agree within the very precise statistical estimates 
with simulation data over the entire fluid density range. An interesting question that arises is whether or not is it
possible to obtain also improved expressions for the free energy of a homogeneous hard-sphere fluid and to derive a corresponding  density functional for the inhomogeneous hard-sphere fluid based on the requirement of a high-power term (in $\eta$) in $\gamma$ and $\Gamma$.

\section{Acknowledgments}
BBL acknowledges support from the National Science Foundation (NSF) under grant CHE-0957102.  The improved simulation data\cite{supp} were obtained using the ALICE High Performance Computing Facility at the University of Leicester.

\newpage

\end{document}